\documentclass[a4paper,aps,twocolumn,nofootinbib]{revtex4}
\RequirePackage[colorlinks,hyperindex]{hyperref}
\RequirePackage[english]{babel}
\RequirePackage[latin1]{inputenc}
\RequirePackage[T1]{fontenc}
\RequirePackage{mathrsfs}
\RequirePackage{amsmath}
\RequirePackage{amssymb}
\RequirePackage{amsbsy}
\RequirePackage{color}
\RequirePackage{bm}
\hypersetup{colorlinks=true,breaklinks=true,urlcolor=blue,linkcolor=red}
\begin{document}
\title{\bf{The Most General Propagator in Quantum Field Theory}}
\author{Luca Fabbri$^{\nabla}$ and Rodolfo Jos\'{e} Bueno Rogerio$^{\hbar}$}
\affiliation{$^{\nabla}$DIME Sez. Metodi e Modelli Matematici, 
Universit\`{a} di Genova, via all'Opera Pia 15, 16145 Genova, ITALY\\
$^{\hbar}$Institute of Physics and Chemistry, Federal 
University of Itajub\'{a}, Itajub\'{a}, Minas Gerais, 37500-903, BRAZIL}
\date{\today}
\begin{abstract}
One of the most important mathematical tools necessary for Quantum Field Theory calculations is the field propagator. Applications are always done in terms of plane waves and although this has furnished many magnificent results, one may still be allowed to wonder what is the form of the most general propagator that can be written. In the present paper, by exploiting what is called polar form, we find the most general propagator in the case of spinors, whether regular or singular, and we give a general discussion in the case of vectors.
\end{abstract}
\maketitle
\section{Introduction}
Quantum Field Theory (QFT) is a set of prescription used to provide some among the most precise predictions in physics. The central mathematical tool needed in QFT is the field propagator. In general, the field propagator is used in one of its simplest forms, that is for plane wave solutions. Nevertheless, one may ask what would happen if more general propagators are used in QFT processes.

In fact, richer propagators might contain in themselves additional information that can be employed for supplementary effects related to radiative corrections. In order to find more general, and even the most general, propagator, we will employ the reduction to the polar form.

The polar form is the process in which spinor fields \cite{G} are written in such a way as to have all components expressed like the product of a module times a phase while still respecting their spinorial covariance. The advantage of this form is that it allows the real degrees of freedom to be isolated from all components related to the frame, therefore rendering the description of the spinor field considerably cleaner. In addition, the components related to the frame can be transferred into the frame itself where they combine with the gauge connection so to give rise to objects that contain the same information of the gauge connection but which can also be proven to be real tensors \cite{Fabbri:2018crr,Fabbri:2019kfr}, with the consequence that this formalism will always display manifest general covariance, and as such it is ready to host any enlargement to curved space-times.

Spinor fields have a very specific structure that can be split into regular and singular \cite{L,Cavalcanti:2014wia,Fabbri:2016msm}. Writing spinors in polar form depends on the specific structure of the spinor field itself, whether regular or singular \cite{Fabbri:2016laz,Fabbri:2020elt}, and consequently the propagator could also feel this discrepancy so that a double analysis should be carried forward.

Additionally, we may also ask what happens not only when spinors, but also when vectors are considered \cite{Fabbri:2019oxy}.

In the present paper we will provide an exhaustive account of the most general propagators in all these cases.
\section{Fundamental Geometry}\label{fundamentalgeometry}
We begin this paper with a section setting the stage.

\subsection{Spinors}

As a start, we introduce the Clifford matrices $\boldsymbol{\gamma}^{a}$ such that $\left\{\boldsymbol{\gamma}_{a},\!\boldsymbol{\gamma}_{b}\right\}\!=\!2\eta_{ab}\mathbb{I}$ with $\eta_{ab}$ being the usual Minkowski matrix. With $\left[\boldsymbol{\gamma}_{a},\!\boldsymbol{\gamma}_{b}\right]\!=\!4\boldsymbol{\sigma}_{ab}$ we define the generators of the complex Lorentz algebra. Then $2i\boldsymbol{\sigma}_{ab}\!=\!\varepsilon_{abcd}\boldsymbol{\pi}\boldsymbol{\sigma}^{cd}$ is the implicit definition of the $\boldsymbol{\pi}$ matrix, which is usually denoted as a gamma matrix with an index five, but since in space-time this index has no meaning, and sometimes it may also be misleading, we prefer to use a notation in which no index appears. An algebra of Clifford matrices also verifies the following general relationships
\begin{eqnarray}
&\boldsymbol{\gamma}_{i}\boldsymbol{\gamma}_{j}\boldsymbol{\gamma}_{k}
\!=\!\boldsymbol{\gamma}_{i}\eta_{jk}-\boldsymbol{\gamma}_{j}\eta_{ik}
\!+\!\boldsymbol{\gamma}_{k}\eta_{ij}
\!+\!i\varepsilon_{ijkq}\boldsymbol{\pi}\boldsymbol{\gamma}^{q}
\end{eqnarray}
are valid as general geometric identities. By exponentiating the generators of the complex Lorentz algebra $\boldsymbol{\sigma}_{ab}$ for local parameters $\theta_{ij}\!=\!-\theta_{ji}$ it is possible to find the local complex Lorentz group $\boldsymbol{S}$ and a spinor field $\psi$ is defined as what transforms as $\psi\!\rightarrow\!\boldsymbol{S}\psi$ in general. With Clifford matrices it is also possible to build a procedure that will convert a spinor $\psi$ in its adjoint spinor $\overline{\psi}\!=\!\psi^{\dagger}\boldsymbol{\gamma}^{0}$ so that we have $\overline{\psi}\!\rightarrow\!\overline{\psi}\boldsymbol{S}^{-1}$ as adjoint transformation. Therefore, with all gamma matrices, and the pair of adjoint spinors, it is possible to construct the bi-linear spinor quantities
\begin{eqnarray}
&\Sigma^{ab}\!=\!2\overline{\psi}\boldsymbol{\sigma}^{ab}\boldsymbol{\pi}\psi\\
&M^{ab}\!=\!2i\overline{\psi}\boldsymbol{\sigma}^{ab}\psi
\end{eqnarray}
\begin{eqnarray}
&S^{a}\!=\!\overline{\psi}\boldsymbol{\gamma}^{a}\boldsymbol{\pi}\psi\\
&U^{a}\!=\!\overline{\psi}\boldsymbol{\gamma}^{a}\psi
\end{eqnarray}
\begin{eqnarray}
&\Theta\!=\!i\overline{\psi}\boldsymbol{\pi}\psi\\
&\Phi\!=\!\overline{\psi}\psi
\end{eqnarray}
which are all real tensors. Nonetheless, these six bi-linear spinor quantities are not independent, since
\begin{eqnarray}
\nonumber
&\psi\overline{\psi}\!\equiv\!\frac{1}{4}\Phi\mathbb{I}
\!+\!\frac{1}{4}U_{a}\boldsymbol{\gamma}^{a}
\!+\!\frac{i}{8}M_{ab}\boldsymbol{\sigma}^{ab}-\\
&-\frac{1}{8}\Sigma_{ab}\boldsymbol{\sigma}^{ab}\boldsymbol{\pi}
\!-\!\frac{1}{4}S_{a}\boldsymbol{\gamma}^{a}\boldsymbol{\pi}
\!-\!\frac{i}{4}\Theta \boldsymbol{\pi}\label{Fierz}
\end{eqnarray}
and then
\begin{eqnarray}
&\Sigma^{ab}\!=\!-\frac{1}{2}\varepsilon^{abij}M_{ij}\\
&M^{ab}\!=\!\frac{1}{2}\varepsilon^{abij}\Sigma_{ij}
\end{eqnarray}
with
\begin{eqnarray}
&M_{ab}\Phi\!-\!\Sigma_{ab}\Theta\!=\!U^{j}S^{k}\varepsilon_{jkab}\label{A1}\\
&M_{ab}\Theta\!+\!\Sigma_{ab}\Phi\!=\!U_{[a}S_{b]}\label{A2}
\end{eqnarray}
and with the orthogonality relations
\begin{eqnarray}
&\frac{1}{2}M_{ab}M^{ab}\!=\!-\frac{1}{2}\Sigma_{ab}\Sigma^{ab}\!=\!\Phi^{2}\!-\!\Theta^{2}
\label{norm2}\\
&\frac{1}{2}M_{ab}\Sigma^{ab}\!=\!-2\Theta\Phi
\label{orthogonal2}\\
&U_{a}U^{a}\!=\!-S_{a}S^{a}\!=\!\Theta^{2}\!+\!\Phi^{2}\label{norm1}\\
&U_{a}S^{a}\!=\!0\label{orthogonal1}
\end{eqnarray}
called Fierz identities. We notice that while the definition of the six bi-linear spinor quantities makes up for a symmetric form of the Fierz identities, nevertheless we could drop either $\Sigma^{ab}$ or $M^{ab}$ since they are the Hodge dual of each other. To conclude the introduction of kinematic quantities, we need to say that by employing the metric we define the symmetric connection as usual with $\Lambda^{\sigma}_{\alpha\nu}$ and with it we define the spin connection according to $\Omega^{a}_{\phantom{a}b\pi}\!=\!\xi^{\nu}_{b}\xi^{a}_{\sigma}(\Lambda^{\sigma}_{\nu\pi}\!-\!\xi^{\sigma}_{i}\partial_{\pi}\xi_{\nu}^{i})$ so that with the gauge potential $qA_{\mu}$ we can finally define
\begin{eqnarray}
&\boldsymbol{\Omega}_{\mu}
=\frac{1}{2}\Omega^{ab}_{\phantom{ab}\mu}\boldsymbol{\sigma}_{ab}
\!+\!iqA_{\mu}\boldsymbol{\mathbb{I}}\label{spinorialconnection}
\end{eqnarray}
called spinorial connection. It is needed to write
\begin{eqnarray}
&\boldsymbol{\nabla}_{\mu}\psi\!=\!\partial_{\mu}\psi
\!+\!\boldsymbol{\Omega}_{\mu}\psi\label{spincovder}
\end{eqnarray}
as spinor covariant derivative. The commutator can justify the definitions of space-time and gauge tensors
\begin{eqnarray}
&R^{i}_{\phantom{i}j\mu\nu}\!=\!\partial_{\mu}\Omega^{i}_{\phantom{i}j\nu}
\!-\!\partial_{\nu}\Omega^{i}_{\phantom{i}j\mu}
\!+\!\Omega^{i}_{\phantom{i}k\mu}\Omega^{k}_{\phantom{k}j\nu}
\!-\!\Omega^{i}_{\phantom{i}k\nu}\Omega^{k}_{\phantom{k}j\mu}\\
&F_{\mu\nu}\!=\!\partial_{\mu}A_{\nu}\!-\!\partial_{\nu}A_{\mu}
\end{eqnarray}
that is the Riemann curvature and the Maxwell strength.

For the dynamics, we take the spinor field subject to
\begin{eqnarray}
&i\boldsymbol{\gamma}^{\mu}\boldsymbol{\nabla}_{\mu}\psi
\!-\!XW_{\mu}\boldsymbol{\gamma}^{\mu}\boldsymbol{\pi}\psi\!-\!m\psi\!=\!0
\label{D}
\end{eqnarray}
in which $W_{\mu}$ is the axial-vector torsion whereas $X$ is the torsion-spin coupling constant, and this is what is called Dirac equation. If we multiply (\ref{D}) by $\boldsymbol{\gamma}^{a}$ and $\boldsymbol{\gamma}^{a}\boldsymbol{\pi}$ and then by $\overline{\psi}$ splitting real and imaginary parts gives
\begin{eqnarray}
\nonumber
&i(\overline{\psi}\boldsymbol{\nabla}^{\alpha}\psi
\!-\!\boldsymbol{\nabla}^{\alpha}\overline{\psi}\psi)
\!-\!\nabla_{\mu}M^{\mu\alpha}-\\
&-XW_{\sigma}M_{\mu\nu}\varepsilon^{\mu\nu\sigma\alpha}\!-\!2mU^{\alpha}\!=\!0\\
\nonumber
&\nabla_{\alpha}\Phi
\!-\!2(\overline{\psi}\boldsymbol{\sigma}_{\mu\alpha}\!\boldsymbol{\nabla}^{\mu}\psi
\!-\!\boldsymbol{\nabla}^{\mu}\overline{\psi}\boldsymbol{\sigma}_{\mu\alpha}\psi)+\\
&+2X\Theta W_{\alpha}\!=\!0\label{vi}
\end{eqnarray}
\begin{eqnarray}
\nonumber
&\nabla_{\nu}\Theta\!-\!
2i(\overline{\psi}\boldsymbol{\sigma}_{\mu\nu}\boldsymbol{\pi}\boldsymbol{\nabla}^{\mu}\psi\!-\!
\boldsymbol{\nabla}^{\mu}\overline{\psi}\boldsymbol{\sigma}_{\mu\nu}\boldsymbol{\pi}\psi)-\\
&-2X\Phi W_{\nu}\!+\!2mS_{\nu}\!=\!0\label{ar}\\
\nonumber
&(\boldsymbol{\nabla}_{\alpha}\overline{\psi}\boldsymbol{\pi}\psi
\!-\!\overline{\psi}\boldsymbol{\pi}\boldsymbol{\nabla}_{\alpha}\psi)
\!-\!\frac{1}{2}\nabla^{\mu}M^{\rho\sigma}\varepsilon_{\rho\sigma\mu\alpha}+\\
&+2XW^{\mu}M_{\mu\alpha}\!=\!0
\end{eqnarray}
as easy to see, and called Gordon decompositions. These are not all the possible decompositions of the Dirac equations, but it will be possible to prove that these four, and in fact a sub-set of these, are enough to account for all information contained in the original Dirac equation itself and therefore nothing else is essentially needed.
\subsection{Vectors}

As for the case of vector fields, the geometrical background requires no additional structure other than the one we have provided in the case of spinors above.

The dynamics is set by assigning the field equations
\begin{eqnarray}
&\nabla_{\sigma}(\partial V)^{\sigma\mu}\!+\!M^{2}V^{\mu}\!=\!\Gamma^{\mu}\label{F}
\end{eqnarray}
with $(\partial V)_{\sigma\mu}\!=\!\partial_{\sigma}V_{\mu}\!-\!\partial_{\mu}V_{\sigma}$ and $\Gamma^{\mu}$ is an external source.
\section{Polar Forms}
What we have presented so far is the general theory, which we are next going to write in term of polar forms.

\subsection{Spinors}
Spinor fields as presented above are characterized by a set of bi-linear quantities that provide the ground for a specific spinor field classification \cite{L,Cavalcanti:2014wia,Fabbri:2016msm, HoffdaSilva:2017waf,daSilva:2012wp,Cavalcanti:2020obq,Ablamowicz:2014rpa,daRocha:2008we, Meert:2018qzk,Rogerio:2019xcu,Rogerio:2020ewe,Rodrigues:2005yz,R1,R2}. We split the cases obtained either when at least one between $\Theta$ or $\Phi$ is not identically zero and giving rise to the \emph{regular} spinors or when $\Theta\!=\!\Phi\!\equiv\!0$ giving rise the \emph{singular} spinors.

\subsubsection{Regular Spinors}
Regular spinors are defined when at least $\Theta\!\neq\!0$ or $\Phi\!\neq\!0$ identically. As an example, they contain Dirac spinors.

In this case (\ref{norm1}) tells that $U^{a}$ is time-like, so that we can always perform up to three boosts to bring its spatial components to vanish. Then, it is always possible to use the rotation around the first and second axes to bring the space part of $S^{a}$ aligned with the third axis. And finally, it is always possible to employ the last rotation to bring the spinor, in chiral representation, into the form
\begin{eqnarray}
&\!\!\psi\!=\!\phi e^{-\frac{i}{2}\beta\boldsymbol{\pi}}
\boldsymbol{S}\left(\!\begin{tabular}{c}
$1$\\
$0$\\
$1$\\
$0$
\end{tabular}\!\right)
\label{regular}
\end{eqnarray}
where $\boldsymbol{S}$ is a general spinor transformation with $\phi$ and $\beta$ called module and Yvon-Takabayashi angle respectively, and this is called polar form of regular spinors \cite{Fabbri:2016msm}. In the polar form regular spinor fields are such that
\begin{eqnarray}
&\Sigma^{ab}\!=\!2\phi^{2}(\cos{\beta}u^{[a}s^{b]}\!-\!\sin{\beta}u_{j}s_{k}\varepsilon^{jkab})\\
&M^{ab}\!=\!2\phi^{2}(\cos{\beta}u_{j}s_{k}\varepsilon^{jkab}\!+\!\sin{\beta}u^{[a}s^{b]})
\end{eqnarray}
showing that the antisymmetric tensors are written in terms of the vectors
\begin{eqnarray}
&S^{a}\!=\!2\phi^{2}s^{a}\\
&U^{a}\!=\!2\phi^{2}u^{a}
\end{eqnarray}
and the scalars
\begin{eqnarray}
&\Theta\!=\!2\phi^{2}\sin{\beta}\\
&\Phi\!=\!2\phi^{2}\cos{\beta}
\end{eqnarray}
also showing that $\beta$ and $\phi$ are a pseudo-scalar and a scalar and thus they are the only real degrees of freedom of the spinor field. Then
\begin{eqnarray}
&\!\!\!\!\psi\overline{\psi}\!\equiv\!\frac{1}{2}
\phi^{2}[(u_{a}\boldsymbol{\mathbb{I}}\!+\!s_{a}\boldsymbol{\pi})\boldsymbol{\gamma}^{a}
\!\!+\!e^{-i\beta\boldsymbol{\pi}}(\boldsymbol{\mathbb{I}}
\!-\!2u_{a}s_{b}\boldsymbol{\sigma}^{ab}\boldsymbol{\pi})]
\end{eqnarray}
and all others become trivial except for the ones given by
\begin{eqnarray}
&u_{a}u^{a}\!=\!-s_{a}s^{a}\!=\!1\\
&u_{a}s^{a}\!=\!0
\end{eqnarray}
showing that the normalized velocity vector $u^{a}$ and spin axial-vector $s^{a}$ are not free. The advantage of writing the spinor fields in polar form is that the $8$ real components are rearranged into the special configuration in which the $2$ real scalar degrees of freedom remain isolated from the $6$ real components that are always transferable into the frame. In general we can formally write that
\begin{eqnarray}
&\boldsymbol{S}\partial_{\mu}\boldsymbol{S}^{-1}\!=\!i\partial_{\mu}\alpha\mathbb{I}
\!+\!\frac{1}{2}\partial_{\mu}\theta_{ij}\boldsymbol{\sigma}^{ij}\label{spintrans}
\end{eqnarray}
so that we can define
\begin{eqnarray}
&\partial_{\mu}\alpha\!-\!qA_{\mu}\!\equiv\!P_{\mu}\label{P}\\
&\partial_{\mu}\theta_{ij}\!-\!\Omega_{ij\mu}\!\equiv\!R_{ij\mu}\label{R}
\end{eqnarray}
which can be proven to be real tensors. Phase and parameters do not alter the information within gauge potential and spin connection, but the non-physical components of spinors encoded by phase and parameters combine with the non-covariant properties of gauge potential and spin connection in order to ensure covariance of $P_{\mu}$ and $R_{ij\mu}$ eventually. For this reason, expressions (\ref{P}, \ref{R}) are called gauge-invariant vector momentum and tensorial connection. With them we can write
\begin{eqnarray}
&\!\!\!\!\!\!\!\!\boldsymbol{\nabla}_{\mu}\psi\!=\!(-\frac{i}{2}\nabla_{\mu}\beta\boldsymbol{\pi}
\!+\!\nabla_{\mu}\ln{\phi}\mathbb{I}
\!-\!iP_{\mu}\mathbb{I}\!-\!\frac{1}{2}R_{ij\mu}\boldsymbol{\sigma}^{ij})\psi
\label{decspinder}
\end{eqnarray}
is the spinorial covariant derivative. We also have
\begin{eqnarray}
&\nabla_{\mu}s_{i}\!=\!R_{ji\mu}s^{j}\label{ds}\\
&\nabla_{\mu}u_{i}\!=\!R_{ji\mu}u^{j}\label{du}
\end{eqnarray}
are general geometric identities. Taking the commutator gives the expressions
\begin{eqnarray}
\!\!\!\!&qF_{\mu\nu}\!=\!-(\nabla_{\mu}P_{\nu}\!-\!\nabla_{\nu}P_{\mu})\label{Maxwell}\\
&\!\!\!\!\!\!\!\!R^{i}_{\phantom{i}j\mu\nu}\!=\!-(\nabla_{\mu}R^{i}_{\phantom{i}j\nu}
\!-\!\!\nabla_{\nu}R^{i}_{\phantom{i}j\mu}
\!\!+\!R^{i}_{\phantom{i}k\mu}R^{k}_{\phantom{k}j\nu}
\!-\!R^{i}_{\phantom{i}k\nu}R^{k}_{\phantom{k}j\mu})\label{Riemann}
\end{eqnarray}
in terms of the Maxwell strength and Riemann curvature, and so they encode electrodynamic and gravitational information, filtering out all information about gauge and frames. However we recall that the gauge-invariant vector momentum and the tensorial connection encode information about electrodynamics and gravity as well as about gauge and frames \cite{Fabbri:2018crr}. Because we can find non-zero solutions of equations (\ref{Maxwell}, \ref{Riemann}) after setting the Maxwell strength and Riemann curvature to zero \cite{Fabbri:2019kfr}, these represent the gauge-invariant vector momentum and the tensorial connection encoding information related to gauge and frames but \emph{not} to electrodynamics and gravity. The resulting gauge-invariant vector momentum and tensorial connection would hence describe a gauge-invariant vector potential as well as a covariant inertial acceleration.

Finally, by substituting the polar form of spinorial covariant derivative of regular spinor fields into the Gordon decompositions and setting
\begin{eqnarray}
&\frac{1}{2}\varepsilon_{\mu\alpha\nu\iota}R^{\alpha\nu\iota}\!=\!B_{\mu}\\
&R_{\mu a}^{\phantom{\mu a}a}\!=\!R_{\mu}
\end{eqnarray}
one can isolate the pair of independent field equations
\begin{eqnarray}
&B_{\mu}\!-\!2P^{\iota}u_{[\iota}s_{\mu]}
\!-\!2XW_{\mu}\!+\!\nabla_{\mu}\beta\!+\!2s_{\mu}m\cos{\beta}\!=\!0\label{dep1}\\
&R_{\mu}\!-\!2P^{\rho}u^{\nu}s^{\alpha}\varepsilon_{\mu\rho\nu\alpha}
\!+\!2s_{\mu}m\sin{\beta}\!+\!\nabla_{\mu}\ln{\phi^{2}}\!=\!0\label{dep2}
\end{eqnarray}
and as it is possible to see they imply (\ref{D}), so that (\ref{dep1}, \ref{dep2}) are equivalent to the initial Dirac spinorial field equations \cite{Fabbri:2016laz}. The spinorial field equations (\ref{D}) are $4$ complex field equations, or $8$ real field equations, which are as many as the $2$ vector field equations given by (\ref{dep1}, \ref{dep2}). Such pair of vector field equations specify all space-time derivatives of both degrees of freedom. The Yvon-Takabayashi angle, in its being the phase difference between chiral projections, has to be related to the mass term, as it is here.

Before concluding this section it is necessary to remark that with the Dirac equations in polar form it is easy to express the gauge-invariant vector momentum in an explicit way. Starting from (\ref{dep1}, \ref{dep2}) and setting
\begin{eqnarray}
&\frac{1}{2}(\nabla_{k}\beta\!-\!2XW_{k}\!+\!B_{k})\!=\!Y_{k}\\
&-\frac{1}{2}(\nabla_{k}\ln{\phi^{2}}\!+\!R_{k})\!=\!Z_{k}
\end{eqnarray}
as well as 
\begin{eqnarray}
&m\cos{\beta}\!-\!Y\!\cdot\!s\!=\!M
\end{eqnarray}
one can prove that
\begin{eqnarray}
&P^{\mu}\!=\!Mu^{\mu}\!+\!Y\!\cdot\!u s^{\mu}\!+\!Z_{k}s_{j}u_{i}\varepsilon^{kji\mu}
\label{momentum}
\end{eqnarray}
showing that the momentum has a component along the velocity but also a component along the spin and a component orthogonal to both velocity and spin \cite{Fabbri:2019tad}.

\subsubsection{Singular Spinors}
Singular spinors are defined by $\Theta\!=\!\Phi\!=\!0$ identically, and they are also known generally as \emph{flag-dipole} spinorial fields. They contain two sub-classes according to whether $S^{a}\!\equiv\!0$ giving \emph{flagpole} spinors, or Majorana spinors, while $M^{ab}\!\equiv\!0$ gives \emph{dipole} spinors, or Weyl spinors, as known.

In this case (\ref{norm2}, \ref{orthogonal2}) tell that if $M_{ab}$ is written in terms of $M_{0K}\!=\!E_{K}$ and $M_{IJ}\!=\!\varepsilon_{IJK}B^{K}$ then they are such that $E^{2}\!=\!B^{2}$ and $\vec{E}\!\cdot\!\vec{B}\!=\!0$ and as above we can always boost and rotate them so to bring $\vec{B}$ and $\vec{E}$ aligned with the first and second axis. Then, the spinor field is
\begin{eqnarray}
&\!\!\psi\!=\!\frac{1}{\sqrt{2}}(\mathbb{I}\cos{\frac{\alpha}{2}}
\!-\!\boldsymbol{\pi}\sin{\frac{\alpha}{2}})\boldsymbol{S}\left(\!\begin{tabular}{c}
$1$\\
$0$\\
$0$\\
$1$
\end{tabular}\!\right)
\label{singular}
\end{eqnarray}
with $\boldsymbol{S}$ a general spin transformation and this is called polar form of singular spinor fields. The case $\alpha\!=\!0$ gives
\begin{eqnarray}
&\!\!\psi\!=\!\frac{1}{\sqrt{2}}\boldsymbol{S}\left(\!\begin{tabular}{c}
$1$\\
$0$\\
$0$\\
$1$
\end{tabular}\!\right)
\end{eqnarray}
such that $\boldsymbol{\gamma}^{2}\psi^{\ast}\!=\!\eta\psi$ with $\eta$ a constant unitary phase and which is the well known case of the Majorana spinorial field, while $\alpha\!=\!\pm\pi/2$ gives
\begin{eqnarray}
&\!\!\psi\!=\!\frac{1}{2}(\mathbb{I}\mp\boldsymbol{\pi})\boldsymbol{S}\left(\!\begin{tabular}{c}
$1$\\
$0$\\
$0$\\
$1$
\end{tabular}\!\right)
\end{eqnarray}
which are the cases of Weyl spinorial fields. In polar form singular spinors verify the relationships
\begin{eqnarray}
&S^{a}\!=\!-\sin{\alpha}U^{a}\label{m}
\end{eqnarray}
showing that the pseudo-scalar $\alpha$ is the only real degree of freedom. The case
\begin{eqnarray}
&S^{a}\!=\!0
\end{eqnarray}
is that of Majorana spinors, the case 
\begin{eqnarray}
&S^{a}\!=\!\mp U^{a}
\end{eqnarray}
as well as
\begin{eqnarray}
&M_{ab}\!=\!0
\end{eqnarray} 
is that of Weyl spinors. Consequently
\begin{eqnarray}
&\psi\overline{\psi}\!\equiv\!\frac{1}{4}U_{a}\boldsymbol{\gamma}^{a}
\!+\!\frac{i}{8}M_{ab}\boldsymbol{\sigma}^{ab}
\!-\!\frac{1}{8}\Sigma_{ab}\boldsymbol{\sigma}^{ab}\boldsymbol{\pi}
\!-\!\frac{1}{4}S_{a}\boldsymbol{\gamma}^{a}\boldsymbol{\pi}
\end{eqnarray}
with 
\begin{eqnarray}
&U_{a}U^{a}\!=\!M_{ab}M^{ab}\!=\!\varepsilon^{abij}M_{ab}M_{ij}\!=\!0\\
&M_{ik}U^{i}\!=\!\varepsilon_{ikab}M^{ab}U^{i}\!=\!0
\end{eqnarray}
identically. Again, the case
\begin{eqnarray}
&\psi\overline{\psi}\!\equiv\!\frac{1}{4}U_{a}\boldsymbol{\gamma}^{a}
\!+\!\frac{i}{4}M_{ab}\boldsymbol{\sigma}^{ab}
\end{eqnarray}
is valid for Majorana, while
\begin{eqnarray}
&\psi\overline{\psi}\!\equiv\!\frac{1}{4}U_{a}\boldsymbol{\gamma}^{a}
(\mathbb{I}\!\pm\!\boldsymbol{\pi})
\end{eqnarray}
is valid for Weyl. With singular spinor fields in polar form we can still define the gauge-invariant vector momentum and tensorial connection as above. For singular spinorial fields we have that
\begin{eqnarray}
\nonumber
&\boldsymbol{\nabla}_{\mu}\psi\!=\![-\frac{1}{2}(\mathbb{I}\tan{\alpha}
\!+\!\boldsymbol{\pi}\sec{\alpha})\nabla_{\mu}\alpha-\\
&-iP_{\mu}\mathbb{I}\!-\!\frac{1}{2}R_{ij\mu}\boldsymbol{\sigma}^{ij}]\psi
\end{eqnarray}
as spinorial covariant derivative. Then
\begin{eqnarray}
&\!\!\nabla_{\mu}U_{i}\!=\!R_{ji\mu}U^{j}\\
&\!\!\!\!\nabla_{\mu}M^{ab}\!=\!-M^{ab}\tan{\alpha}\nabla_{\mu}\alpha
\!-\!R^{a}_{\phantom{a}k\mu}M^{kb}\!+\!R^{b}_{\phantom{b}k\mu}M^{ka}
\end{eqnarray}
are valid as geometric identities. Of course, the commutator would give the same results as for regular spinors.

Finally, by substituting the polar form of spinorial covariant derivative of singular spinor fields into the Gordon decompositions, then plugging the bi-linear spinorial quantities, and diagonalizing the results, we get
\begin{eqnarray}
&\!\!(\varepsilon^{\mu\rho\sigma\nu}\nabla_{\mu}\alpha\sec{\alpha}
\!-\!2P^{[\rho}g^{\sigma]\nu})M_{\rho\sigma}\!=\!0\\
&\!\!\!\!M_{\rho\sigma}(g^{\nu[\rho}\nabla^{\sigma]}\alpha\sec{\alpha}
\!-\!2P_{\mu}\varepsilon^{\mu\rho\sigma\nu})\!+\!4m\sin{\alpha}U^{\nu}\!=\!0
\end{eqnarray}
\begin{eqnarray}
\nonumber
&\!\![(2XW\!-\!B)^{\sigma}\varepsilon_{\sigma\mu\rho\nu}\!+\!R_{[\mu}g_{\rho]\nu}+\\
&+g_{\nu[\mu}\nabla_{\rho]}\alpha\tan{\alpha}]M_{\eta\zeta}\varepsilon^{\mu\rho\eta\zeta}\!=\!0\\
\nonumber
&\!\!\!\![(2XW\!-\!B)^{\sigma}\varepsilon_{\sigma\mu\rho\nu}\!+\!R_{[\mu}g_{\rho]\nu}+\\
&+g_{\nu[\mu}\nabla_{\rho]}\alpha\tan{\alpha}]M^{\mu\rho}\!+\!4mU_{\nu}\!=\!0
\end{eqnarray}
equivalent to the Dirac spinorial field equations \cite{Fabbri:2020elt}. They specify all derivatives of the degree of freedom as should be expected. The cases of Majorana and Weyl spinorial fields however have no remaining degree of freedom.

Some solution to these equations have first been found \cite{Vignolo:2011qt,daRocha:2013qhu} and we believe that with the polar form the quest for further solutions might be generally simplified.
\subsection{Vectors}
Compared to spinors, the case of vectors is simple.

If the vector is time-like then with the same derivation done above we can prove that we can always write
\begin{eqnarray}
&V\!=\!\phi\Lambda\left(\!\begin{tabular}{c}
$1$\\
$0$\\
$0$\\
$0$
\end{tabular}\!\right)
\label{pos}
\end{eqnarray}
while for light-like vectors we can always write 
\begin{eqnarray}
&V\!=\!\Lambda\left(\!\begin{tabular}{c}
$1$\\
$0$\\
$0$\\
$1$
\end{tabular}\!\right)
\label{nul}
\end{eqnarray}
and for space-like vectors we can always write
\begin{eqnarray}
&V\!=\!\phi\Lambda\left(\!\begin{tabular}{c}
$0$\\
$0$\\
$0$\\
$1$
\end{tabular}\!\right)
\label{neg}
\end{eqnarray}
with $\Lambda$ a general real Lorentz transformation and $\phi$ called module, and this is called polar form of vectors. In the following, it will be simpler to define
\begin{eqnarray}
&V^{a}\!=\!\phi v^{a}
\end{eqnarray}
in terms of a $v^{b}$ such that $v^{2}\!=\!1$, $v^{2}\!=\!0$ or $v^{2}\!=\!-1$ for the three cases above, and showing that the module is a scalar and consequently the only degree of freedom.

In any of these cases we have that we can write
\begin{eqnarray}
&(\Lambda)^{i}_{\phantom{i}k}\partial_{\mu}(\Lambda^{-1})^{k}_{\phantom{k}j}
\!=\!\partial_{\mu}\theta^{i}_{\phantom{i}j}
\end{eqnarray}
so that we can define
\begin{eqnarray}
&\partial_{\mu}\theta^{i}_{\phantom{i}j}\!-\!\Omega^{i}_{\phantom{i}j\mu}
\!\equiv\!R^{i}_{\phantom{i}j\mu}
\end{eqnarray}
which is the tensorial connection as above. With it
\begin{eqnarray}
&\nabla_{\mu}V^{a}\!=\!(\delta^{a}_{b}\nabla_{\mu}\ln{\phi}\!-\!R^{a}_{\phantom{a}b\mu})V^{b}
\label{decvecder}
\end{eqnarray}
as covariant derivative. Then we have
\begin{eqnarray}
&\nabla_{\mu}v_{a}\!=\!R_{ba\mu}v^{b}\label{dv}
\end{eqnarray}
as general identities. The commutator still gives the same expression for the Riemann curvature in general.

When in (\ref{F}) we plug the polar form we get
\begin{eqnarray}
\nonumber
&(g^{\alpha\nu}\nabla^{2}\phi\!-\!\nabla^{\nu}\nabla^{\alpha}\phi-\\
\nonumber
&-R^{\nu}\nabla^{\alpha}\phi\!+\!R^{\nu\alpha\sigma}\nabla_{\sigma}\phi
\!+\!R^{\nu[\alpha\sigma]}\nabla_{\sigma}\phi+\\
&+\nabla_{\sigma}R^{\nu[\alpha\sigma]}\phi
\!+\!R^{\sigma[\alpha\pi]}R^{\nu}_{\phantom{\nu}\sigma\pi}\phi
\!+\!M^{2}g^{\alpha\nu}\phi)v_{\nu}\!=\!\Gamma^{\alpha}
\end{eqnarray}
as the polar form of the field equations for vector fields.

The interested reader can have a look at reference \cite{Fabbri:2019oxy}.

With this long but necessary introduction, we have all we need to find the propagators in the most general cases.
\section{General Propagators}
We are now ready to employ the polar form to invert the field equations and obtain the propagators.

We split again the cases of spinors and vectors.

\subsection{Spinors}
In the case of the Dirac spinor field equations (\ref{D}), to establish the propagator we write the Dirac spinorial field equation with (\ref{decspinder}) in polar form and introduce
\begin{eqnarray}
&\frac{1}{2}(\nabla_{k}\beta\!-\!2XW_{k}\!+\!B_{k})\!=\!Y_{k}\\
&-\frac{1}{2}(\nabla_{k}\ln{\phi^{2}}\!+\!R_{k})\!=\!Z_{k}
\end{eqnarray}
as well as 
\begin{eqnarray}
&E_{k}\!=\!Y_{k}\\
&F_{k}\!=\!P_{k}\!-\!iZ_{k}
\end{eqnarray}
so that
\begin{eqnarray}
&(F_{k}\boldsymbol{\gamma}^{k}\!+\!E_{k}\boldsymbol{\gamma}^{k}\boldsymbol{\pi}\!-\!m)\psi\!=\!0
\label{eqpolar}
\end{eqnarray}
is the expression of the Dirac equation. Then the expression of the propagator $G$ is the solution to the equation
\begin{eqnarray}
&(F_{k}\boldsymbol{\gamma}^{k}\!+\!E_{k}\boldsymbol{\gamma}^{k}\boldsymbol{\pi}\!-\!m)G
\!=\!\mathbb{I}
\end{eqnarray}
which we have to find in the most general circumstance.

To do that, multiply on the left by the general matrix given by $F_{k}\boldsymbol{\gamma}^{k}\!-\!E_{k}\boldsymbol{\gamma}^{k}\boldsymbol{\pi}\!-\!m\mathbb{I}$ so to get
\begin{eqnarray}
\nonumber
&[-2mF_{k}\boldsymbol{\gamma}^{k}\!+\!2F\!\cdot\!E\boldsymbol{\pi}
\!+\!(m^{2}\!+\!F^{2}\!+\!E^{2})\mathbb{I}]G=\\
&=F_{k}\boldsymbol{\gamma}^{k}\!-\!E_{k}\boldsymbol{\gamma}^{k}\boldsymbol{\pi}\!-\!m\mathbb{I}
\end{eqnarray}
and then again by $2mF_{k}\boldsymbol{\gamma}^{k}\!-\!2F\!\cdot\!E\boldsymbol{\pi}\!+\!(F^{2}\!+\!E^{2}\!+\!m^{2})\mathbb{I}$ as
\begin{eqnarray}
\nonumber
&\left[(F^{2}\!+\!E^{2}\!+\!m^{2})^{2}\!-\!4m^{2}F^{2}\!-\!|2F\!\cdot\!E|^{2}\right]G=\\
\nonumber
&=[2mF_{i}\boldsymbol{\gamma}^{i}\!-\!2F\!\cdot\!E\boldsymbol{\pi}
\!+\!(F^{2}\!+\!E^{2}\!+\!m^{2})\mathbb{I}]\cdot\\
&\cdot(F_{k}\boldsymbol{\gamma}^{k}\!-\!E_{k}\boldsymbol{\gamma}^{k}\boldsymbol{\pi}
\!-\!m\mathbb{I})
\end{eqnarray}
which can now be easily inverted. In fact we have
\begin{eqnarray}
\nonumber
&G\!=\!\left[(F^{2}\!+\!E^{2}\!+\!m^{2})^{2}\!-\!4m^{2}F^{2}
\!-\!|2F\!\cdot\!E|^{2}\right]^{-1}\cdot\\
\nonumber
&\cdot[2mF_{i}\boldsymbol{\gamma}^{i}\!-\!2F\!\cdot\!E\boldsymbol{\pi}
\!+\!(F^{2}\!+\!E^{2}\!+\!m^{2})\mathbb{I}]\cdot\\
&\cdot(F_{k}\boldsymbol{\gamma}^{k}\!-\!E_{k}\boldsymbol{\gamma}^{k}\boldsymbol{\pi}
\!-\!m\mathbb{I})\label{G}
\end{eqnarray}
as the most general propagator for regular spinors.

Similarly we may set in the case of singular spinors
\begin{eqnarray}
&\frac{1}{2}(B_{k}\!-\!2XW_{k})\!=\!Y_{k}\label{1}\\
&-\frac{1}{2}(R_{k}\!-\!\tan{\alpha}\nabla_{k}\alpha)\!=\!Z_{k}\label{2}
\end{eqnarray}
as well as 
\begin{eqnarray}
&E_{k}\!=\!Y_{k}\!-\!i\sec{\alpha}\nabla_{k}\alpha/2\label{3}\\
&F_{k}\!=\!P_{k}\!-\!iZ_{k}\label{4}
\end{eqnarray}
so that
\begin{eqnarray}
&(F_{k}\boldsymbol{\gamma}^{k}\!+\!E_{k}\boldsymbol{\gamma}^{k}\boldsymbol{\pi}\!-\!m)\psi\!=\!0
\end{eqnarray}
exactly as in the situation above.

Therefore, we have the same form for the most general propagator in the case of singular spinors.
\subsection{Vectors}
In the case of the vector field equations (\ref{F}), we might substitute (\ref{decvecder}) and introduce the matrix
\begin{eqnarray}
\nonumber
&g^{\alpha\nu}\nabla^{2}\phi\!-\!\nabla^{\nu}\nabla^{\alpha}\phi-\\
\nonumber
&-R^{\nu}\nabla^{\alpha}\phi\!+\!R^{\nu\alpha\sigma}\nabla_{\sigma}\phi
\!+\!R^{\nu[\alpha\sigma]}\nabla_{\sigma}\phi+\\
&+\nabla_{\sigma}R^{\nu[\alpha\sigma]}\phi
\!+\!R^{\sigma[\alpha\pi]}R^{\nu}_{\phantom{\nu}\sigma\pi}\phi
\!+\!M^{2}g^{\alpha\nu}\phi\!=\!A^{\alpha\nu}
\end{eqnarray}
so to have
\begin{eqnarray}
&A^{\alpha\nu}v_{\nu}\!=\!\Gamma^{\alpha}
\end{eqnarray}
as the expression for the vector field equations. The propagator is the object $G_{\nu\rho}$ such that 
\begin{eqnarray}
&A^{\alpha\nu}G_{\nu\rho}\!=\!\delta^{\alpha}_{\rho}
\end{eqnarray}
and which we have to find in general.

Here the matrix we have to invert is not constituted in terms of gamma matrices, whose algebraic properties are well known, and because we have no specific information on the structure of the $A^{\alpha\nu}$ matrix, then its inversion is generally not possible, at least with the same method.

We shall now discuss how the spinor propagator does in fact contain the propagator of the standard QFT.
\section{Perturbative QFT}
The restriction to QFT is done by imposing the condition of plane waves on the spinors. This means that
\begin{eqnarray}
&i\boldsymbol{\nabla}_{\mu}\psi\!=\!P_{\mu}\psi
\end{eqnarray}
and upon comparison with the most general form given by expression (\ref{decspinder}) we obtain that
\begin{eqnarray}
&(\nabla_{\mu}\ln{\phi}\mathbb{I}
\!-\!\frac{i}{2}\nabla_{\mu}\beta\boldsymbol{\pi}
\!-\!\frac{1}{2}R_{ij\mu}\boldsymbol{\sigma}^{ij})\psi=0
\end{eqnarray}
for any spinor field. Then since $\boldsymbol{\sigma}^{ij}$, $\mathbb{I}$ and $\boldsymbol{\pi}$ are linearly independent we get $R_{ij\mu}=0$ with $\beta$ and $\phi$ constant. For constant pseudo-scalars the only possible value is zero, and thus we get $R_{ij\mu}=0$ with $\beta\!=\!0$ and $\nabla_{\nu}\phi\!=\!0$ as the conditions that implement the reduction to plane waves and hence the reduction to the environment of QFT.

When these restrictions are implemented, and also torsion is neglected, we get that
\begin{eqnarray}
&Y_{k}\!=\!0\\
&Z_{k}\!=\!0
\end{eqnarray}
so that 
\begin{eqnarray}
&M\!=\!m
\end{eqnarray}
and (\ref{momentum}) reduces to
\begin{eqnarray}
&P^{\mu}\!=\!mu^{\mu}
\label{momentumreduced}
\end{eqnarray}
as is usual in QFT. The same approximation can be obtained when $\beta\!=\!0$ also for $s^{a}\!\rightarrow\!0$ and that is in the case of spinless approximation. Remark that while the common belief wants the momentum to be expressed in terms of the velocity alone, we have proved here that this can be done only when the internal structures are neglected, as in the macroscopic limit. In general, any internal dynamics would forbid non-relativistic regimes even in the rest frame, and thus (\ref{momentumreduced}) cannot be valid in such cases.

With the same restriction we obtain
\begin{eqnarray}
&E_{k}\!=\!0\\
&F_{k}\!=\!P_{k}
\end{eqnarray}
and so
\begin{eqnarray}
\nonumber
&G\!=\!\left[(P^{2}\!+\!m^{2})^{2}\!-\!4m^{2}P^{2}\right]^{-1}\cdot\\
&\cdot[2mP_{i}\boldsymbol{\gamma}^{i}\!+\!(P^{2}\!+\!m^{2})\mathbb{I}]
\!\cdot\!(P_{k}\boldsymbol{\gamma}^{k}\!-\!m\mathbb{I})
\end{eqnarray}
and more specifically
\begin{eqnarray}
&G\!=\!(P^{2}\!-\!m^{2})^{-1}\cdot(P_{k}\boldsymbol{\gamma}^{k}\!+\!m\mathbb{I})
\label{propagatorspinless}
\end{eqnarray}
which is precisely the usual propagator of QFT. As such it is singular on-shell. Because the same expression for the propagator can also be obtained in the spinless approximation, then it is possible to give a partial interpretation of QFT singular behaviour at the poles, and that is the singularity comes from neglecting the spin, and more in general the internal structure. This is to be expected for point-like particles such as those commonly used in QFT.

Consequently, we can say that the QFT propagator is the simplest of all possible propagators that can be employed to calculate scattering processes in general.

It may now be interesting to see what happens when QFT conditions are not taken exactly. For such a purpose we may relax the conditions $R_{ij\mu}=0$, $\beta\!=\!0$ and $\nabla_{\nu}\phi\!=\!0$, or $s^{a}\!=\!0$, and take them as small corrections to QFT. In this way, we have that $E_{k}$ could be taken to be negligible compared to $F_{k}$ and in $F_{k}$ we can take $Z_{k}$ small compared to $P_{\mu}$ in the propagator. Then we have
\begin{eqnarray}
\nonumber
&\!\!\!\!G\!=\!(P^{2}\!-\!m^{2}\!-\!2iP\!\cdot\!Z)^{-2}
\!\cdot\![(P^{2}\!-\!m^{2})(P_{k}\boldsymbol{\gamma}^{k}\!+\!m\mathbb{I})-\\
\nonumber
&\!\!\!\!-2iP\!\cdot\!Z(P_{k}\boldsymbol{\gamma}^{k}\!+\!m\mathbb{I})
\!-\!(P^{2}\!-\!m^{2})iZ_{k}\boldsymbol{\gamma}^{k}-\\
&\!\!\!\!-2mP_{[i}E_{k]}\boldsymbol{\sigma}^{ik}\boldsymbol{\pi}
\!-\![(P^{2}\!+\!m^{2})E_{k}\!-\!2P\!\cdot\!EP_{k}]\boldsymbol{\gamma}^{k}\boldsymbol{\pi}]
\end{eqnarray}
to first order in $E_{k}$ and $Z_{k}$ alike. Remark that the term $-2iP\!\cdot\!Z$ in the denominator plays the role of the $i\epsilon$ added by hand in QFT. Terms in $Z_{k}$ quite generally are related to the presence of a finite size for field distributions. With no such contribution we would have 
\begin{eqnarray}
\nonumber
&\!\!G\!=\!(P^{2}\!-\!m^{2})^{-2}
\!\cdot\![(P^{2}\!-\!m^{2})(P_{k}\boldsymbol{\gamma}^{k}\!+\!m\mathbb{I})+\\
&\!\!+2mi\varepsilon^{ikab}P_{i}E_{k}\boldsymbol{\sigma}_{ab}
\!-\!2(P^{2}E_{k}\!-\!P\!\cdot\!EP_{k})\boldsymbol{\gamma}^{k}\boldsymbol{\pi}]
\end{eqnarray}
having used the $P^{2}\!=\!m^{2}$ approximation. Notice that the terms in $\boldsymbol{\sigma}_{ab}$ and $\boldsymbol{\gamma}^{k}\boldsymbol{\pi}$ indicate the effects of spin contributions. Terms in $E_{k}$ generally are linked to the internal structure. And without such an internal structure we are going to reduce to the usual propagator of QFT.

It is therefore clear how the most general propagator can in fact be seen as some correction around the simplest propagator of QFT, with the correction being constituted of two terms encoded by $Z_{k}$ when the matter distribution has a size and by $E_{k}$ when there is an internal structure.
\section{Non-Perturbative QFT}
There is a point that is now essential to high-light, and it concerns the fact that in the case of the singular spinor the general propagator contains a mass term. Mass terms in this case may be allowed in general for flag-dipoles, and in particular, with some adjustments, for flag-poles, that is the Majorana case. However, for di-poles, that is Weyl spinors, the mass term should be set to zero. This creates problems for the inversion of the Dirac operator.

In perturbative QFT, Weyl spinors are treated by using both chiralities since the chiral information is contained inside the vertices \cite{DeWitt}. Nevertheless, in non-perturbative QFT, where the full propagator is necessary, this can no longer be done. To see how this issue can be resolved in non-perturbative QFT let us consider it explicitly.

As we have said, the problems arise in the case of Weyl spinors, after taking the limit $m\!\rightarrow\!0$ in the exact propagator (\ref{G}). The result is then
\begin{eqnarray}
\nonumber
&G\!=\!\left[(F^{2}\!+\!E^{2})^{2}\!-\!|2F\!\cdot\!E|^{2}\right]^{-1}\cdot\\
&\cdot[-2F\!\cdot\!E\boldsymbol{\pi}\!+\!(F^{2}\!+\!E^{2})\mathbb{I}]
\!\cdot\!(F_{k}\boldsymbol{\gamma}^{k}\!-\!E_{k}\boldsymbol{\gamma}^{k}\boldsymbol{\pi})
\end{eqnarray}
where we recall that
\begin{eqnarray}
&E_{k}\!=\!Y_{k}\!-\!i\sec{\alpha}\nabla_{k}\alpha/2\\
&F_{k}\!=\!P_{k}\!-\!iZ_{k}
\end{eqnarray}
with
\begin{eqnarray}
&\frac{1}{2}(B_{k}\!-\!2XW_{k})\!=\!Y_{k}\\
&-\frac{1}{2}(R_{k}\!-\!\tan{\alpha}\nabla_{k}\alpha)\!=\!Z_{k}
\end{eqnarray}
because of (\ref{3}, \ref{4}) and (\ref{1}, \ref{2}). Weyl spinors are defined by the fact that $\alpha$ is a constant, which means that
\begin{eqnarray}
&E_{k}\!=\!\frac{1}{2}(B_{k}\!-\!2XW_{k})\\
&F_{k}\!=\!P_{k}\!+\!\frac{i}{2}R_{k}
\end{eqnarray}
so if torsion is negligible and the tensorial connection is zero, as in QFT, we get $E_{k}=0$ and $P_{k}\!=\!F_{k}$ so that
\begin{eqnarray}
&G\!=\!(P^{2})^{-1}P_{k}\boldsymbol{\gamma}^{k}
\end{eqnarray}
and because for Weyl in QFT it is $P^2=0$ then we have a singular point and therefore lack of invertibility.

Nevertheless, this is true only in standard QFT, where plane waves are used. However, this is not the case when a more general treatment of QFT is considered. In fact, when torsion is not negligible or the tensorial connection is not zero, the general Weyl propagator is
\begin{eqnarray}
\nonumber
&G\!=\!\left[(F^{2}\!+\!E^{2})^{2}\!-\!|2F\!\cdot\!E|^{2}\right]^{-1}\cdot\\
&\cdot[-2F\!\cdot\!E\boldsymbol{\pi}\!+\!(F^{2}\!+\!E^{2})\mathbb{I}]
\cdot(F_{k}\boldsymbol{\gamma}^{k}\!-\!E_{k}\boldsymbol{\gamma}^{k}\boldsymbol{\pi})
\end{eqnarray}
which would be singular if we were to have a zero at the denominator, and thus
\begin{eqnarray}
&(F^{2}\!+\!E^{2})^{2}\!-\!|2F\!\cdot\!E|^{2}\!=\!0
\end{eqnarray}
which can be re-written as
\begin{eqnarray}
&F^{2}\!+\!E^{2}\!\pm\!2F\!\cdot\!E\!=\!0
\end{eqnarray}
or equivalently
\begin{eqnarray}
&(F^{a}\!\pm\!E^{a})(F_{a}\!\pm\!E_{a})=0
\end{eqnarray}
in general. As a consequence
\begin{eqnarray}
&B_{k}\!-\!2XW_{k}\!\pm\!2P_{k}\!=\!0\\
&R_{k}\!=\!0
\end{eqnarray}
which are relationships between fields that are in principle independent and so they can not be generally verified.

What this means is that in general situations there can be no singularity and the propagator will be invertible in the case in which $m\!=\!0$ as for the Weyl spinor.

This circumstance also helps us see something important for general treatments of QFT, that is QFT might have problems of singularity coming from the fact that point-particles are themselves singular, but all these problems might not arise in a more general context.

The assumption of plane wave expansion that is at the basis of QFT is certainly a powerful hypothesis for simplification but it may very well carry with it too strong approximations generating problems that would not even appear if the study is done in more general environments.
\section{Feynman-Dyson Analysis}\label{quantumfields}
In this section we introduce one of the most important mathematical tools for the treatment of interacting fields with perturbative methods, that is the \emph{Feynman-Dyson propagator}. We start by defining quantum fields as
\begin{eqnarray}
\mathfrak{F}(x)\!=\!\int\frac{d^3 P}{(2\pi)^3}\frac{1}{\sqrt{2E(\boldsymbol{P})}}\sum_{spin}\Big[c_{\iota}(\boldsymbol{P})\psi_{\iota}(\boldsymbol{P}) e^{-iP_{\mu}x^{\mu}}
\nonumber\\
+d_{\iota}^{\dag}(\boldsymbol{P})\chi_{\iota}(\boldsymbol{P}) e^{iP_{\mu}x^{\mu}}\Big]
\end{eqnarray}
and with the adjoint spinor $\bar{\psi}$ we have the adjoint
\begin{eqnarray}
\bar{\mathfrak{F}}(x)\!=\!\int\frac{d^3 P}{(2\pi)^3}\frac{1}{\sqrt{2E(\boldsymbol{P})}}\sum_{spin}\Big[c_{\iota}^{\dag}(\boldsymbol{P})\bar{\psi}_{\iota}(\boldsymbol{P}) e^{iP_{\mu}x^{\mu}}
\nonumber\\
+d_{\iota}(\boldsymbol{P})\bar{\chi}_{\iota}(\boldsymbol{P}) e^{-iP_{\mu}x^{\mu}}\Big]
\end{eqnarray}
in which $\chi$ stands for the antiparticle spinor. Creation and annihilation operators $c^{\dag}(\boldsymbol{P})$ and $c(\boldsymbol{P})$ yield
\begin{eqnarray}
&&\{c_{\iota}(\boldsymbol{P}),c_{\iota^{\prime}}^{\prime\dag}(\boldsymbol{P}^{\prime})\}\!=\!(2\pi)^3\delta^{3}(\boldsymbol{P}-\boldsymbol{P}^{\prime})\delta_{\iota\iota^{\prime}}
\\
&&\{d_{\iota}(\boldsymbol{P}),d_{\iota^{\prime}}^{\prime\dag}(\boldsymbol{P}^{\prime})\}\!=\!(2\pi)^3\delta^{3}(\boldsymbol{P}-\boldsymbol{P}^{\prime})\delta_{\iota\iota^{\prime}}
\\
&&\{c_{\iota}(\boldsymbol{P}),d_{\iota^{\prime}}(\boldsymbol{P}^{\prime})\}\!=\!\{c^{\dag}_{\iota}(\boldsymbol{P}),d^{\dag}_{\iota^{\prime}}(\boldsymbol{P}^{\prime})\}\!=\!0
\end{eqnarray}  
as it is usual in QFT. The Feynman-Dyson propagator is given by the time-ordered product of $\mathfrak{F}(x)$ and $\bar{\mathfrak{F}}(x)$ as
\begin{eqnarray}\label{fdpropagator}
i\mathcal{D}(x-x^{\prime})&\!=\!&\langle 0 \vert \mathfrak{F}(x)\bar{\mathfrak{F}}(x^{\prime}) \vert 0 \rangle  \theta(x^{0}-x^{\prime 0})\nonumber\\
&-& \langle 0 \vert \bar{\mathfrak{F}}(x^{\prime})\mathfrak{F}(x) \vert 0 \rangle \theta(x^{\prime 0}-x^{0})
\end{eqnarray}
in terms of the Heaviside function $\theta(t)$ in Fourier representation. Having defined the quantum fields operators, now we proceed by evaluating the two terms in the FD propagator. The first term of \eqref{fdpropagator} is given by
\begin{eqnarray}\label{fd1}
&&\langle 0\vert \mathfrak{F}(x)\bar{\mathfrak{F}}(x^{\prime}) \vert 0 \rangle\theta(x^{0}-x^{\prime 0})\!=\!-\frac{1}{2\pi i}\int\frac{d^3 P}{(2\pi)^3}\frac{1}{2E(\boldsymbol{P})}
\nonumber
\\
&&\times \int_{-\infty}^{\infty}ds\sum_{spin}\psi_{\iota}(\boldsymbol{P})\bar{\psi}_{\iota}(\boldsymbol{P})\frac{e^{-i(s+P_{0})(x^{0}-x^{\prime 0})+i\boldsymbol{P}(\boldsymbol{x}-\boldsymbol{x}^{\prime})}}{s+i\epsilon}
\nonumber
\\
\end{eqnarray}
for $x^0> x^{\prime0}$ while for the evaluation of the second term \newpage
\begin{eqnarray}\label{fd2}
&&\langle 0 \vert \bar{\mathfrak{F}}(x^{\prime}){\mathfrak{F}}(x) \vert 0 \rangle \theta(x^{\prime 0}-x^{0})\!=\!-\frac{1}{2\pi i}\int\frac{d^3 P}{(2\pi)^3}\frac{1}{2E(\boldsymbol{P})}
\nonumber
\\
&&\times\int_{-\infty}^{\infty}ds\sum_{spin}\chi_{\iota}(\boldsymbol{P})\bar{\chi}_{\iota}(\boldsymbol{P})\frac{e^{i(s+P_{0})(x^{0}-x^{\prime 0})-i\boldsymbol{P}(\boldsymbol{x}-\boldsymbol{x}^{\prime})}}{s+i\epsilon}
\nonumber
\\
\end{eqnarray}
for $x^0< x^{\prime0}$. Plugging \eqref{fd1} and \eqref{fd2} in \eqref{fdpropagator} and using the Heaviside function properties, one reaches the following form for the Feynman-Dyson propagator
\begin{widetext} 
\begin{eqnarray}\label{fd3}
&&\mathcal{D}(x-x^{\prime})\!=\!\int\frac{d^4 P}{(2\pi)^4}\frac{1}{2E(\boldsymbol{P})}\Bigg[\frac{\sum_{spin}\psi_{\iota}(\boldsymbol{P})\bar{\psi}_{\iota}(\boldsymbol{P})(P_0+\sqrt{P_{j}P^{j}+m^2})}{P^{2}-m^2+i\epsilon}\nonumber\\
&+&\frac{\sum_{spin}\chi_{\iota}(-\boldsymbol{P})\bar{\chi}_{\iota}(-\boldsymbol{P})(P_0-\sqrt{P_{j}P^{j}+m^2})}{P^2-m^2+i\epsilon} \Bigg]e^{-iP_{\mu}(x^{\mu}-x^{\prime\mu})}\nonumber\\
\end{eqnarray}
\end{widetext}
with $j=1, 2, 3$. Inserting the spin-sum expression in \eqref{fd3}, and caring to write all expression in manifestly covariant form we finally obtain that
\begin{widetext}
\begin{eqnarray}\label{fdfinal}
&&\mathcal{D}(x-x^{\prime})\!=\!\int\frac{d^4 P}{(2\pi)^4}\frac{[\gamma_{\mu}(Mu^{\mu}+Y\!\cdot\!us^{\mu}+Z_{k}s_{j}u_{i}\epsilon^{kji\mu})+m\mathbb{I}]}{P^{2}-m^2+i\epsilon}e^{-iP_{\mu}(x^{\mu}-x^{\prime\mu})}
\end{eqnarray}
\end{widetext}
as the Feynman-Dyson propagator with spinors in polar form. The spinless counterpart of \eqref{fdfinal} is
\begin{eqnarray}\label{fdspinless}
&&\!\!\!\!\!\!\!\!\!\!\!\!\!\!\!\!\mathcal{D}(x-x^{\prime})\!=\!\int\frac{d^4 P}{(2\pi)^4}\frac{(P_{\mu}\gamma^{\mu}+m\mathbb{I})}{P^{2}-m^2+i\epsilon}e^{-iP_{\mu}(x^{\mu}-x^{\prime\mu})}
\end{eqnarray}
as it is easy to see.

The propagators we presented in \eqref{G} and \eqref{propagatorspinless} are the core of the propagators defined in \eqref{fdfinal} and \eqref{fdspinless}, and they are intrinsically related by a Fourier transformation.

In the elementary particle physics framework, the most common type of event is related to scattering processes, in which the cross-section is defined. The computation of such a quantity is accomplished via the propagation amplitude. Indeed the Feynman propagator will turn out to be part of the Feynman rules, the expressions above will attach to internal lines of the Feynman diagrams, representing, then, the particle propagation. When perturbative calculations are taken into account with Feynman diagrams, commonly one associate the \eqref{G} and \eqref{propagatorspinless}, or \eqref{fdfinal} and \eqref{fdspinless}, with each of the internal fermionic lines.
\section{Conclusion}
In this article, we have considered the polar form of vectors, as well as singular and regular spinors, to search for the most general propagator in quantum field theory.

We have seen that while for vectors the use of the polar form does not seem to bring any advantage, for both singular and regular spinors the polar form allows a re-writing of the field equations that permits the full inversion and therefore the most general propagator is found.

We discussed how this spinorial general propagator does in fact contain the one usually employed in QFT but it also contains two types of corrections, one coming from the fact that the matter distribution has a finite size and one from the fact that it has an internal structure.

We eventually demonstrated how these results can be tied to those usually obtained by exploiting standard QFT calculations as the Feynman-Dyson propagator.

Having thus established the form of the most general propagator that can be used in QFT, what remains to be done consists in finding possible applications for specific scattering processes, and hope that the additional information contained in this general propagator may give supplementary corrections, capable of being observed.

\end{document}